\documentclass[10pt]{article}

\usepackage{graphicx}
\usepackage[T1]{fontenc}
\usepackage[utf8]{inputenc}
\usepackage{amssymb}
\usepackage{amsfonts}
\usepackage{dsfont}
\usepackage{mathtools}
\usepackage{amsthm}
\usepackage{amsmath}
\usepackage{yhmath}
\usepackage{relsize}
\usepackage{textcomp}
\usepackage{eurosym}
\usepackage{stmaryrd}
\usepackage{xcolor}
\usepackage[multiple]{footmisc}
\usepackage{bigints}
\usepackage{geometry}
\newtheorem{prop}{Proposition}

\newtheorem{rem}{Remark}

\begin{document}

\title{To Hedge or Not to Hedge: Optimal Strategies for Stochastic Trade Flow Management}

\author{Philippe \textsc{Bergault}\footnote{Université Paris Dauphine-PSL, Ceremade, Paris, France, bergault@ceremade.dauphine.fr.} \and Hamza \textsc{Bodor}\footnote{Université Paris 1 Panthéon-Sorbonne, UFR 27 Mathématiques et Informatique, Centre d'Economie de la Sorbonne, Paris, France, and BNP Paribas Corporate and Institutional Banking, Global Markets Data \& Artificial
Intelligence Lab, Paris, France, hamza.bodor@bnpparibas.com.}  \and Olivier \textsc{Guéant}\footnote{Université Paris 1 Panthéon-Sorbonne, UFR 27 Mathématiques et Informatique, Centre d'Economie de la Sorbonne, Paris, France, olivier.gueant@univ-paris1.fr.}}
\date{}

\maketitle
\setlength\parindent{0pt}

\begin{abstract}

This paper addresses the trade-off between internalisation and externalisation in the management of stochastic trade flows. We consider agents who must absorb flows and manage risk by deciding whether to warehouse it or hedge in the market, thereby incurring transaction costs and market impact. Unlike market makers, these agents cannot skew their quotes to attract offsetting flows and deter risk-increasing ones, leading to a fundamentally different problem. Within the Almgren-Chriss framework, we derive almost-closed-form solutions in the case of quadratic execution costs, while more general cases require numerical methods. In particular, we discuss the challenges posed by artificial boundary conditions when using classical grid-based numerical PDE techniques and propose reinforcement learning methods as an alternative.

\vspace{5mm}

\end{abstract}

\section{Introduction}

Since the notable renewal of academic research topics following the 2008 financial crisis, market impact has assumed a central role in the quantitative finance literature. Over the past two decades, numerous datasets have been analysed to estimate the market impact of (i) individual limit and market orders (see for instance \cite{bouchaud2018trades}, \cite{cont2014price} and \cite{eisler2012price}), and (ii) meta-orders (see for example \cite{almgren_direct_2005}, \cite{bucci_crossover_2019} and \cite{moro_market_2009}). Beyond empirical and phenomenological studies documenting the so-called square-root law(s) of market impact and other stylised facts, microfoundations and theoretical justifications have been proposed to enhance our understanding of the microscopic dynamics of prices and limit order books (see for instance \cite{benzaquen_market_2018}, \cite{bucci_co-impact_2020} and \cite{donier_fully_2015}) and delineate the types of models suitable for practical applications (see \cite{gatheral_no-dynamic-arbitrage_2010} for early discussions on the topic, particularly the necessary linearity of permanent market impact).\\

Overall, a variety of market impact models have been proposed: highly realistic models with nonlinear and / or non-Markovian transient dynamics have been developed and applied to specific problems, while simpler models, mainly inspired by the Almgren-Chriss \cite{almgren_optimal_2001, almgren_optimal_2003} and Obizhaeva-Wang \cite{obizhaeva2013optimal} frameworks, have been utilised to ensure the tractability of equations within otherwise complex models.\\

Market impact models have indeed served as foundational components in the development of optimal execution strategies, proving useful for brokers optimising their arrival price / implementation shortfall or VWAP algorithms (e.g. \cite{alfonsi_optimal_2010}, \cite{almgren_optimal_2001}, \cite{almgren_optimal_2003}, \cite{cartea_algorithmic_2015}, \cite{frei2015optimal}, \cite{gueant2016financial}, \cite{gueant2014vwap} and \cite{obizhaeva2013optimal}), as well as for algorithmic traders aiming to exploit alpha signals without excessively impacting the market (see for instance \cite{fouque_optimal_2022}, \cite{hey2023cost}, \cite{hey_trading_2023} and \cite{lehalle_incorporating_2019}). Moreover, they have demonstrated their value to help designing strategies to better hedge large positions in derivatives or contingent claims written on assets with high execution costs (see \cite{almgren_option_2016}, \cite{gueant_option_2017} and \cite{rogers_cost_2010}).\\

More recently, market impact models have been used to develop frameworks addressing the financial trade-off between internalising and externalising trading flows (see \cite{barzykin2022dealing}, \cite{barzykin2022market}, \cite{barzykin2023algorithmic} -- see also \cite{butz_internalisation_2019} and \cite{cartea2022brokers} for related discussions).\footnote{Pre-hedging is a related topic on which a few papers have recently been written, see for instance \cite{muhle2024pre}.} In a nutshell, given a stochastic flow of buy and sell trades received in a book, to what extent is it optimal to keep the risk in the book and wait for offsetting flows, rather than hedging the risk directly on the market, thereby incurring transaction costs and price impact? This issue is, of course, faced by market makers and dealers whose activity -- across all OTC markets -- involves quoting buy and sell prices to their clients and using their balance sheet to warehouse the risk until an opposite order arrives, or until they decide to utilise their outside option of actively trading with other dealers on the inter-dealer segment of the market (see \cite{schrimpf2019fx} for a discussion on the two segments of the FX market).\\

Unlike market makers, who have significant flexibility to propose skewed prices that attract or deter one side of the flow in order to rebalance their positions (or even to stop quoting a product on the bid or ask side entirely), certain participants -- for example, those dealing with retail equity flows or managing central risk books -- must absorb flows and manage the associated risk appropriately. Their choices lie in deciding whether to continue bearing the risk or, alternatively, to offload it on markets while optimising the trading speed to mitigate execution costs.\\

Although practices have evolved with the advent of competition between trading platforms associated with MiFID and Reg. NMS, which has encouraged various participants to establish dark pools and crossing networks to offer parts of their inventories to other participants while limiting market impact, there exists, to our knowledge, only one model, presented in \cite{nutz2023unwinding}, addressing the trade-off between internalisation and externalisation beyond the case of market makers.\\

The authors of \cite{nutz2023unwinding}, motivated by the development of central risk books in many banks, developed a model where the flow of trades is stochastic and can follow Brownian, positively autocorrelated, or mean-reverting dynamics. Their model is, in fact, an extension of optimal execution models \textit{à la} Obizhaeva-Wang with no risk aversion, where the inventory is not fixed but rather stochastic. In particular, they model transient price impact using a linear push factor and resilience at an exponential rate. Additionally, they consider execution costs, but in quadratic form only. Because of their specific choice of transient market impact and quadratic execution costs, the model reduces to a linear-quadratic (LQ) optimal control problem that can be solved in closed form using the solution of a Riccati differential equation.\\

The aim of this article is to propose different models for optimising the management of a stochastic trade flow in the risk-averse case, using the Almgren-Chriss framework for execution costs and market impact. More precisely, we consider a risk-averse trader whose objective is either to maximise the expected CARA utility or a risk-adjusted expectation (\textit{à la} Cartea-Jaimungal; see \cite{cartea_algorithmic_2015}) of their profit at a terminal time~$T$. We show that, in both cases, the problem with quadratic execution costs can be tackled almost explicitly, as it reduces to solving a matrix Riccati differential equation. However, for other forms of execution costs, the problem becomes more challenging due to the uncontrolled nature of trade flows, which results in Hamilton-Jacobi(-Bellman) equations that cannot be solved in closed form and may be difficult to approximate numerically because of their sensitivity to artificially-imposed boundary conditions (except in the limiting case of linear bid-ask spread costs).\\ 

In Section 2, we present the models and the associated partial differential equations. In Section 3, we show that these equations can be solved in closed form in the case of quadratic execution costs, up to solving a matrix Riccati differential equation. In Section 4, we address numerical resolution. In particular, we demonstrate how the limiting case of linear execution costs can be tackled using classical techniques, while arguing that reinforcement learning techniques are viable alternatives to classical grid-based numerical PDE techniques for addressing more general cases.

\section{Internalisation vs. Externalisation in the Almgren-Chriss Model}

\subsection{Mathematical Framework}

\subsubsection{State Variables}

We focus on a single asset and a trader managing a book that receives stochastic trade flows over the time interval $[0,T]$, where $T > 0$.\\

Throughout this paper, we consider a filtered probability space $\left(\Omega, \mathcal F, \mathbb P; \mathbb F = (\mathcal F_t)_{t \in [0,T]} \right)$ satisfying the usual conditions. This probability space is assumed to be sufficiently large to support all the processes introduced. By default, all processes are adapted to the above filtration.\\

Assuming Brownian dynamics for the trade flows,\footnote{This framework can be generalised to include trade flows with positive or negative autocorrelation, as in \cite{nutz2023unwinding}.} and absolutely continuous trading strategies, the inventory process $(q_t)_{t \in [0,T]}$ evolves as:
\begin{align}
\label{q}
    dq_t = v_t dt + \nu dB_t,
\end{align}
where $q_0 \in \mathbb R$ and $\nu \in \mathbb R_+$ are given, $(v_t)_{t \in [0,T]}$ represents the trading velocity of the trader, and $(B_t)_{t \in [0,T]}$ is a standard Brownian motion.\\

The asset price process $(S_t)_{t \in [0,T]}$ is modelled as a Brownian motion impacted by the trader's strategy:
\begin{align}
\label{S}
    dS_t = k v_t dt + \sigma dW_t,
\end{align}
where $S_0 \in \mathbb{R}$ is given, $\sigma \in \mathbb R^*_+$ is the volatility parameter, $k \in \mathbb R_+$ represents the permanent market impact parameter, and $(W_t)_{t \in [0,T]}$ is a standard Brownian motion. We assume that, together, $(B_t)_{t \in [0,T]}$ and $(W_t)_{t \in [0,T]}$ form a two-dimensional Brownian motion with a quadratic covariation coefficient $\rho \in [-1,1]$.\\

The trader's cash account $(X_t)_{t \in [0,T]}$ evolves according to:
\begin{align}
\label{X}
    dX_t = - v_t S_t dt - L(v_t) dt,
\end{align}
where $X_0 \in \mathbb R$ is given, and $L: \mathbb R \rightarrow \mathbb R_+$ represents the instantaneous market impact and/or execution costs incurred. In this paper, we primarily focus on cases where $L$ is a strictly convex and asymptotically super-linear function, such as $L(v) = \frac{\psi}{2} |v| + \eta v^{1+\phi}$, with $\psi \in \mathbb R_+$, $\eta \in \mathbb R^*_+$, and $\phi \in \mathbb R^*_+$. The limiting case $\eta \to 0$ will also be discussed.

\subsubsection{Objective Functions}

We consider two optimisation problems. In the first, the trader aims to maximise the expected CARA utility of their profit at terminal time $T$:
\[
\text{PnL}_T := X_T + q_T S_T - \frac{k}{2} q_T^2 - \ell(q_T),
\]
where $\frac{k}{2} q_T^2$ accounts for the cost of liquidation associated with permanent market impact (as in \textit{à la} Almgren-Chriss models), and $\ell(q_T)$ represents additional execution costs at time $T$ (see \cite{gueant2015optimal}). The corresponding optimisation problem is:
\begin{align*}
\sup_{v \in \mathcal A_A} \mathbb E\left[-\exp \left( - \gamma \left(X_T + q_T S_T  - \frac{k}{2} q_T^2 - \ell(q_T) \right)\right) \right], \tag{Model A}
\end{align*}
where $\gamma$ denotes the absolute risk aversion parameter of the trader and where $\mathcal{A}_A$ is the set of admissible controls that consists in adapted processes with additional integrability constraints.\\

Using Eqs.~\eqref{q} and \eqref{S}, and applying Itô's formula, the PnL at time $T$ can be expressed as:
\[
\text{PnL}_T = X_0 + q_0 S_0 + \int_0^T\left(- L(v_t) + kv_tq_t + \rho \nu \sigma\right) dt + \int_0^T \nu S_t dB_t + \int_0^T \sigma q_t dW_t - \frac{k}{2} q_T^2 - \ell(q_T).
\]
To approximate the variance of the PnL, we can follow \textit{à la} Cartea-Jaimungal models and consider only the variance of the stochastic integrals:
\[
\mathbb V\left(\int_0^T \nu S_t dB_t + \int_0^T \sigma q_t dW_t\right) = \int_0^T\left(\nu^2 S_t^2 + \sigma^2 q_t^2 + 2 \rho \nu \sigma q_t S_t\right)dt.
\]
The second optimisation problem we consider, a mean-variance-like formulation, is subsequently:
\begin{align*}
\sup_{v \in \mathcal A_B} \mathbb E\left[\int_0^T\left(- L(v_t) + kv_tq_t + \rho \nu \sigma - \frac{\gamma}{2} \left(\sigma^2 q_t^2 + \nu^2 S_t^2 + 2\rho \sigma \nu q_t S_t \right) \right) dt - \frac{k}{2} q_T^2 - \ell(q_T)\right], \tag{Model B}
\end{align*}
where $\frac \gamma 2$ is the constant penalising the proxy of the variance term and where $\mathcal{A}_B$ is the set of admissible controls defined as
$$\mathcal A_B = \left\{ (v_t)_{t\in [0,T]},\; \mathbb R\text{-valued},\; \mathbb F\text{-adapted} \left| \mathbb E \left[ \int_0^T v_t^2 dt \right] < + \infty \right. \right\}.$$ 

\subsection{Hamilton-Jacobi-Bellman Equations and Changes of Variables}

\subsubsection{Model A}

The Hamilton-Jacobi-Bellman equation associated with Model A is:
\begin{eqnarray}
 0&=& \partial_t u(t,x,q,S) + \frac{1}{2} \sigma^2 \partial^2_{SS}u(t,x,q,S) + \frac{1}{2} \nu^2 \partial^2_{qq}u(t,x,q,S) + \rho \sigma \nu \partial^2_{qS}u(t,x,q,S)\nonumber \\
 &&+\ \underset{v}{\sup} \Big(v \partial_q u(t,x,q,S) - \big(vS + L(v)\big)\partial_xu(t,x,q,S) + kv\partial_S u(t,x,q,S)\Big),
\label{HJBA}
\end{eqnarray}
with terminal condition $u(T,x,q,S)=  -\exp \left( -\gamma \left(x+qS -\frac{k}{2}q^2 - \ell(q) \right)\right)$.\\

Classical results from the optimal execution literature (see \cite{gueant2016financial}) suggest to consider the ansatz
$$ u(t,x,q,S) = -\exp \left( -\gamma \left(x+qS -\frac{k}{2}q^2 + \theta(t,q,S) \right)\right).$$

We then easily see that to obtain a solution to Eq. \eqref{HJBA}, it is sufficient to obtain a solution to
\begin{eqnarray}
0 &=& \!\!\!\!\partial_t \theta(t,q,S)
+ \frac{1}{2}\sigma^2\partial^2_{SS}\theta(t,q,S) +  \frac{1}{2}\nu^2 \partial^2_{qq}\theta(t,q,S) + \rho\sigma\nu \partial^2_{qS}\theta(t,q,S)\nonumber \\
&& \!\!\!\!-\ \frac \gamma 2 \sigma^2 (q + \partial_S \theta(t,q,S))^2
- \frac{\gamma}{2}\nu^2(S - k q + \partial_q \theta(t,q,S))^2 - \gamma\rho\sigma\nu(S - k q + \partial_q \theta(t,q,S))(q + \partial_S \theta(t,q,S))\nonumber \\
&& \!\!\!\!-\ \frac 12 \nu^2 k + \rho \sigma \nu + H\left(\partial_q \theta(t,q,S)+ k\partial_S \theta(t,q,S)\right),
\label{PDE_theta_A}
\end{eqnarray}
where $H(p) = \sup_v v p - L(v)$, and the associated terminal condition is $\theta(T,q,S) = - \ell(q)$.\\

In particular, the candidate optimal control writes
$$v_t^* = H'(\partial_q \theta(t,q_t,S_t)+ k\partial_S \theta(t,q_t,S_t)).$$ 
     
\subsubsection{Model B}

The Hamilton-Jacobi-Bellman equation associated with Model B is:
\begin{eqnarray}
 0&=& \partial_t u(t,q,S) + \frac{1}{2} \sigma^2 \partial^2_{SS}u(t,q,S) + \frac{1}{2} \nu^2 \partial^2_{qq}u(t,q,S) + \rho \sigma \nu \partial^2_{qS}u(t,q,S)\nonumber \\
&& -\ \frac{\gamma}{2} \sigma^2 q^2 - \frac{\gamma}{2} \nu^2 S^2 -\gamma\rho \nu \sigma q S \nonumber \\
&&+\ \rho \sigma \nu + \underset{v}{\sup} \Big(v \partial_q u(t,q,S) - L(v)+ kvq + kv\partial_S u(t,q,S)\Big),
\label{HJBB}
\end{eqnarray}
with terminal condition $u(T,q,S)= -\frac{k}{2}q^2 - \ell(q)$.\\

Here, we use the change of variables $u(t,q,S) = -\frac{k}{2}q^2 + \theta(t,q,S)$ and we easily see that to obtain a solution to Eq. \eqref{HJBB}, it is sufficient to obtain a solution to
\begin{eqnarray}
 0&=& \partial_t \theta(t,q,S) + \frac{1}{2} \sigma^2 \partial^2_{SS}\theta(t,q,S) + \frac{1}{2} \nu^2 \partial^2_{qq}\theta(t,q,S) + \rho \sigma \nu \partial^2_{qS}\theta(t,q,S)\nonumber \\
&& -\ \frac{\gamma}{2} \sigma^2 q^2 - \frac{\gamma}{2} \nu^2 S^2 -\gamma\rho \nu \sigma q S \nonumber \\
&& -\ \frac 12 \nu^2 k + \rho \sigma \nu +  H\left(\partial_q \theta(t,q,S)+ k\partial_S \theta(t,q,S)\right),
\label{PDE_theta_B}
\end{eqnarray}
and the associated terminal condition is $\theta(T,q,S) = - \ell(q)$.\\

In particular, the candidate optimal control writes
$$v_t^* = H'(\partial_q \theta(t,q_t,S_t)+ k\partial_S \theta(t,q_t,S_t)).$$

\section{The Quadratic Execution Cost Case}

To address the internalisation vs. externalisation trade-off, we consider in this section the benchmark case corresponding to quadratic execution costs,\footnote{The case of quadratic costs can easily be generalized to a portfolio of correlated assets.} i.e., $L(v) = \eta v^2$ and $\ell(q) = Kq^2$, where $\eta \in \mathbb{R}_+^*$ and $K \in \mathbb{R}_+$. In this case, the solutions of both Eqs.~\eqref{PDE_theta_A} and \eqref{PDE_theta_B} are (up to a time-dependent additive constant) time-dependent quadratic forms in the variable $\begin{pmatrix} q \\ S \end{pmatrix}$. Furthermore, the matrices of these quadratic forms satisfy matrix Riccati differential equations and can therefore be efficiently computed numerically.\\

Let us start with Model A and the following result, whose proof follows from straightforward computation:
\begin{prop}
Let $\Sigma = \begin{pmatrix}
\sigma^2 & \rho \sigma \nu \\
\rho \sigma \nu & \nu^2
\end{pmatrix}$ and $J = \begin{pmatrix}
0 & 1 \\
1 & 0
\end{pmatrix}$ .\\
Let $$U = \frac{1}{\eta} \begin{pmatrix}
1 & k \\
k & k^2
\end{pmatrix} - 2 \gamma J \Sigma J, \ Y = \gamma J\Sigma \begin{pmatrix}
1 & 0 \\
-k & 1
\end{pmatrix} \ \text{and} \ Q = - \frac \gamma 2 \begin{pmatrix}
1 & 0 \\
-k & 1
\end{pmatrix}^\intercal \Sigma \begin{pmatrix}
1 & 0 \\
-k & 1
\end{pmatrix}.$$
Let $A \in C^1([0,T], S_2(\mathbb{R}))$ be a solution of the matrix Riccati differential equation
\[
A'(t) = A(t) U A(t) + A(t) Y + Y^\intercal A(t) + Q
\]
with the terminal condition 
\[
A(T) = 
\begin{pmatrix}
K & 0 \\
0 & 0
\end{pmatrix}.
\]

Then, $$\theta(t,q,S) = -\begin{pmatrix} q \\ S \end{pmatrix}^\intercal A(t) \begin{pmatrix} q \\ S \end{pmatrix} + \int_t^T \text{Tr}(J\Sigma J A(s)) ds  + \left(\frac 12 \nu^2 k - \rho \sigma \nu\right)(T-t)$$
defines a solution to Eq. \eqref{PDE_theta_A} and satisfies the terminal condition $\theta(T,q,S) = -Kq^2$.\\  
\end{prop}

\begin{rem}
Local existence (and uniqueness) of a solution to the above Riccati equation for $A$ is trivial. However, global existence on $[0,T]$ is not and we have no guarantee of no blow up in finite time for Model A. For this reason, we concentrate in what follows on Model B.
\end{rem}

\begin{rem}
In Model A, using the notations of the above proposition, the candidate optimal control writes
$$v_t^* = - \frac 1 \eta \begin{pmatrix} 1 \\ k \end{pmatrix}^\intercal A(t) \begin{pmatrix} q_t \\ S_t \end{pmatrix}$$
\end{rem}

For Model B we have a simpler result, whose proof also follows from straightforward computation:
\begin{prop}
Let $\Sigma = \begin{pmatrix}
\sigma^2 & \rho \sigma \nu \\
\rho \sigma \nu & \nu^2
\end{pmatrix}$ and $J = \begin{pmatrix}
0 & 1 \\
1 & 0
\end{pmatrix}$ .\\
Let $A \in C^1([0,T], S_2(\mathbb{R}))$ be a solution of the matrix Riccati differential equation
\[
A'(t) = \frac{1}{\eta} A(t) 
\begin{pmatrix}
1 & k \\
k & k^2
\end{pmatrix}
A(t) - \frac \gamma 2 \Sigma,
\]
with the terminal condition 
\[
A(T) = 
\begin{pmatrix}
K & 0 \\
0 & 0
\end{pmatrix}.
\]

Then, $$\theta(t,q,S) = -\begin{pmatrix} q \\ S \end{pmatrix}^\intercal A(t) \begin{pmatrix} q \\ S \end{pmatrix} + \int_t^T \text{Tr}(J\Sigma J A(s)) ds  + \left(\frac 12 \nu^2 k - \rho \sigma \nu\right)(T-t)$$
defines a solution to Eq. \eqref{PDE_theta_B} and satisfies the terminal condition $\theta(T,q,S) = -Kq^2$.\\  
\end{prop}

\begin{rem}
Unlike for Model A, the existence (and uniqueness) on $[0,T]$ of a solution to the above Riccati equation for $A$ is clear.
\end{rem}

\begin{rem}
In Model B, using the notations of the above proposition, the candidate optimal control writes
$$v_t^* = - \frac 1 \eta \begin{pmatrix} 1 \\ k \end{pmatrix}^\intercal A(t) \begin{pmatrix} q_t \\ S_t \end{pmatrix}$$
\end{rem}

\section{Beyond Quadratic Execution Costs}

\subsection{Introductory Remarks on Boundaries}

The numerical approximation of the solution of partial differential equations such as Eqs.~\eqref{PDE_theta_A} and \eqref{PDE_theta_B} requires the introduction of artificial boundaries for both inventories and prices. To better understand the issue with the financial problem at stake, let us focus on Model B in the specific case where $k = \rho = 0$. In this case, the partial differential equation \eqref{PDE_theta_B} simplifies to:
\begin{eqnarray}
 0 &=& \partial_t \theta(t,q,S) + \frac{1}{2} \sigma^2 \partial^2_{SS}\theta(t,q,S) + \frac{1}{2} \nu^2 \partial^2_{qq}\theta(t,q,S) - \frac{\gamma}{2} \sigma^2 q^2 - \frac{\gamma}{2} \nu^2 S^2 + H\left(\partial_q \theta(t,q,S)\right).
\label{PDE_theta_B_0}
\end{eqnarray}

Using the ansatz
\[
\theta(t,q,S) = \tilde{\theta}(t,q) - \alpha(t) - \beta(t) S^2,
\]
solving Eq.~\eqref{PDE_theta_B} boils down to solving
\begin{eqnarray}
 0 &=& \partial_t \tilde{\theta}(t,q) + \frac{1}{2} \nu^2 \partial^2_{qq}\tilde{\theta}(t,q) - \frac{\gamma}{2} \sigma^2 q^2 + H\left(\partial_q \tilde{\theta}(t,q)\right),
\label{PDE_theta_B_0_q}
\end{eqnarray}
with the terminal condition $\tilde{\theta}(T,q) = -\ell(q)$,
along with the system of ordinary differential equations:
\begin{equation*}
\begin{aligned}
\alpha'(t) &= -\sigma^2 \beta(t), \quad \alpha(T) = 0, \\
\beta'(t) &= -\frac{\gamma}{2} \nu^2, \quad \beta(T) = 0,
\end{aligned}
\end{equation*}
whose solution is:
\[
\alpha(t) = \frac{\gamma}{4} \nu^2 \sigma^2 (T-t)^2, \qquad \beta(t) = \frac{\gamma}{2} \nu^2 (T-t).
\]

In the case of quadratic execution costs, and following Proposition 2, $\tilde{\theta}(t,q)$ is, up to a time-dependent additive term independent of $q$, equal to $-a(t) q^2$, where:
\[
a'(t) = -\frac{1}{2} \gamma \sigma^2 + \frac{1}{\eta} a(t)^2, \qquad a(T) = K.
\]
Solving this equation using classical techniques leads to:
\begin{equation}
a(t) = \sqrt{\frac{\gamma \sigma^2 \eta}{2}} \frac{ 1 + \frac{ K - \sqrt{\frac{\gamma \sigma^2 \eta}{2}} }{K + \sqrt{\frac{\gamma \sigma^2 \eta}{2}}}\exp\left(-2\sqrt{\frac{\gamma \sigma^2}{2\eta}} (T-t)\right)}{ 1 - \frac{ K - \sqrt{\frac{\gamma \sigma^2 \eta}{2}} }{K + \sqrt{\frac{\gamma \sigma^2 \eta}{2}}}\exp\left(-2\sqrt{\frac{\gamma \sigma^2}{2\eta}} (T-t)\right)}.
\end{equation}

In particular, we obtain:
\begin{equation}
v^*_t = - \sqrt{\frac{\gamma \sigma^2}{2\eta}} \frac{ 1 + \frac{ K - \sqrt{\frac{\gamma \sigma^2 \eta}{2}} }{K + \sqrt{\frac{\gamma \sigma^2 \eta}{2}}}\exp\left(-2\sqrt{\frac{\gamma \sigma^2}{2\eta}} (T-t)\right)}{ 1 - \frac{ K - \sqrt{\frac{\gamma \sigma^2 \eta}{2}} }{K + \sqrt{\frac{\gamma \sigma^2 \eta}{2}}}\exp\left(-2\sqrt{\frac{\gamma \sigma^2}{2\eta}} (T-t)\right)}q_t.
\label{PDE_psi_0_solution_control}
\end{equation}

This result indicates that the trader trades continuously in proportion to their current inventory -- the proportionality coefficient being time-dependent but independent of $\nu$, as in the standard Almgren-Chriss case with constant (non-random) inventories.\\

To numerically approximate the solution to Eq.~\eqref{PDE_theta_B_0_q} beyond quadratic execution costs, one must artificially impose inventory boundaries at, say, $\pm Q$, where $Q \in \mathbb{R}_+^*$ is large, and choose boundary conditions for $\tilde \theta(t,\cdot)$ at $\pm Q$. In general, selecting the correct boundary condition is impossible (think of dealing with the quadratic case without knowing the closed-form expression) and the error propagates inside the domain. However, there is a limiting case corresponding to linear bid-ask spread costs, i.e., $L(v) = \frac{\psi}{2} |v|$ and $\ell(q) = \frac{\psi}{2} |q|$, for which boundary conditions are natural.\\

In the limiting case of linear execution costs, the problem must in fact be formulated differently: instead of trading continuously, the trader selects a sequence of stopping times $0 \le \tau_1 \le \ldots \le \tau_i \le \ldots$ and corresponding random variables $\xi_1, \ldots, \xi_i, \ldots$ where $\xi_i$ represents the quantity traded at time $\tau_i$. The inventory process is then
\[
q_t = q_0 + \nu B_t + \sum_{\tau_i \le t} \xi_i,
\]
the price follows
\[
S_t = S_0 + \sigma W_t,
\]
and the trader's cash account evolves as
\[
X_t = X_0 - \sum_{\tau_i \le t} \xi_i S_{\tau_i} - \frac{\psi}{2} \sum_{\tau_i \le t} |\xi_i|.
\]

At terminal time $T$, the trader's PnL is:
\[
\text{PnL}_T = X_T + q_T S_T - \frac{\psi}{2} |q_T| = X_0 + q_0 S_0 + \int_0^T \sigma q_{t-} dW_t + \int_0^T \nu S_{t} dB_t  - \frac{\psi}{2} \sum_{\tau_i \le t} |\xi_i| - \frac{\psi}{2} |q_T|.
\]

The limiting case of Model B corresponds therefore to maximising:
\[
\mathbb{E} \left[- \frac{\psi}{2} \sum_{\tau_i \le T} |\xi_i| - \frac{\psi}{2} |q_T| - \frac{\gamma}{2} \int_0^T \sigma^2 q_{t-}^2 dt - \frac{\gamma}{2} \int_0^T \nu^2 S_{t}^2 dt\right]
\]
over the set of stopping times and admissible trade sizes.\\

This is an impulse control problem, leading to the following quasi-variational inequality (QVI):
\[
0 = \max \begin{cases}
    \partial_t \theta(t,q,S) + \frac{1}{2} \sigma^2 \partial^2_{SS}\theta(t,q,S) + \frac{1}{2} \nu^2 \partial^2_{qq}\theta(t,q,S) - \frac{\gamma}{2} \sigma^2 q^2 - \frac{\gamma}{2} \nu^2 S^2, \\
    \underset{\xi}{\max}\, \theta(t, q + \xi) - \theta(t,q) - \frac{\psi}{2} |\xi|
\end{cases}
\]
with terminal condition $\theta(T,q) = - \frac{\psi}{2}|q|$.\\

Using the same ansatz as for Eq. \eqref{PDE_theta_B_0}, this equation reduces to
$$
0 = \max \begin{cases}
    \partial_t \tilde \theta(t,q) + \frac{1}{2} \nu^2 \partial^2_{qq} \tilde\theta(t,q) - \frac{\gamma}{2} \sigma^2 q^2, \\
    \underset{\xi}{\max}\, \tilde\theta(t, q + \xi) - \tilde\theta(t,q) - \frac{\psi}{2} |\xi|
\end{cases}
$$
with terminal condition $\tilde \theta(T,q) = - \frac{\psi}{2}|q|$.\\

The optimal strategy is deduced from the above QVI through the maximiser $\xi$. The QVI has no closed-form solution but can be solved numerically using a grid-based scheme, assuming that $Q$ is chosen sufficiently large such that any trade bringing $|q|$ close to $Q$ is automatically hedged in the market at a cost per asset equal to $\psi/2$. In other words, in the case of linear execution costs, if $Q$ is sufficiently large, the boundary condition should be given by $\partial_q \tilde \theta(t,\pm Q) = \mp \frac{\psi}{2}$. Of course this reasoning cannot generalize to strictly convex execution costs and errors from the boundaries will propagate inside the domain when approximating the solution of PDEs.\\

\subsection{Policy Learning via Reinforcement Learning}

Reinforcement Learning (RL) is a machine learning paradigm in which an agent optimizes a policy through interactions with an environment and feedbacks in the form of rewards. Unlike supervised learning, which relies on labeled data, RL employs trial and error to maximize cumulative rewards. Over the years, RL has demonstrated remarkable success across various domains, including robotics, game playing, and, more recently, finance \cite{gueant2016financial, kolm2019modern}.\\

In finance, RL has gained significant attention for its ability to tackle dynamic and stochastic decision-making problems. Its adaptability allows it to learn nearly-optimal strategies without requiring explicit assumptions about market dynamics, making it particularly valuable in complex environments. RL has been applied to portfolio optimization, market making, high-frequency trading, and optimal execution, where it enables adaptive decision-making that responds to evolving market conditions, even in high dimensions \cite{dixon2020applications, pippas2024evolution}. For a comprehensive review of RL techniques in financial applications, we refer readers to \cite{fischer2018reinforcement}.\\

A key area of application is portfolio management, where RL optimizes asset allocation based on market fluctuations \cite{gueant2019deep, jiang2017deep, liu2018adaptive, wang2021policy, zhang2020meta}. In market making, RL assists in setting bid and ask quotes and managing inventory risk efficiently \cite{chan2001electronic, ganesh2019market, spooner2020adversarial, spooner2018market}. Similarly, in optimal execution, RL helps minimizing execution costs and market impact in dynamic trading environments while, sometimes, minimizing risk \cite{ deng2016deep, lin2020hierarchical, nevmyvaka2006reinforcement, wei2019optimizing}.\\

The traditional way to handle the aforementioned financial problems relies on stochastic optimal control methods, such as Hamilton-Jacobi-Bellman equations and the associated grid-based numerical approximation methods. While these methods are widely used and often lead to very good results, they often require to bound the state space and to artificially impose boundary conditions. Unlike HJB-based approaches, RL is simulation-based and learns directly from the environment's dynamics, eliminating the need for artificial  boundary conditions. This flexibility allows RL to adapt naturally to the properties of the considered environment, making it particularly suitable for settings where specifying relevant artificial boundary conditions is difficult or impractical.\\

In the following sections, we examine how reinforcement learning can be used to manage stochastic order flow while minimizing execution costs and mitigating inventory risk. In particular, we compare the performance of RL-based policies with those obtained from HJB-based numerical methods.\\

\subsection{Numerical Results}

\subsubsection{Experimental Setting}

To evaluate different strategies for managing stochastic trade flow, derived from partial differential equations or reinforcement learning, we set up a fixed parametric market environment.\\

Unless specified otherwise, inventory is measured in lots of 5000 shares.\footnote{This scaling primarily enhances numerical stability in reinforcement learning by normalizing inventory values.}\\

\textbf{Environment parameters:}

\begin{itemize}
    \item Trading horizon: $T = 1\;\text{day}$
    \item Discrete time step: $\Delta t = \frac{1}{100}\;\text{day}$
    \item Initial lot price: $S_0 = 500,000\;\$\cdot\text{lot}^{-1} $
    \item Arithmetic volatility: $\sigma = 10,000\;\$\cdot\text{lot}^{-1}\cdot\text{day}^{-1/2}$
    \item Trade flow standard deviation: $\nu = 10\;\text{lot}\cdot\text{day}^{-1/2}$
    \item Bid-ask spread: $\psi = 250\;\$\cdot\text{lot}^{-1}$
    \item Running speed penalty: $\eta = 5\;\$\cdot\text{lot}^{-2}\cdot\text{day}$
    \item Execution cost exponent: $\phi = 1 ~ (L(v) = \frac{\psi}{2} |v| + \eta v^{2})$
    \item Risk aversion parameter: $\gamma = 2\cdot 10^{-6}\;\$^{-1}$
    \item Terminal inventory penalty: $K = 500\;\$\cdot\text{lot}^{-2}$
\end{itemize}

These parameters characterize the environment we use to define the market dynamics, the evolution of the inventory, and the objective function. In this section on numerical results, we focus on Model B. The objective function in this model balances expected costs with price fluctuations and inventory risk. It also discourages excessively aggressive trading and incentivizes almost complete inventory unwinding by the end of the time interval.\\

In order to approximate the solution of the HJB equation \eqref{HJBB} on $[0,T] \times [-40, 40]$, we use a monotone implicit Euler scheme on a grid of size $161$ for the inventory, with Neumann conditions at the boundaries corresponding to the bid-ask spread costs, i.e. $\partial_q \tilde \theta(t,\pm 40) = \mp \frac{\psi}{2}$.\\

Our reinforcement learning framework is built upon \texttt{mbt-gym} \cite{jerome2023mbt} and employs Proximal Policy Optimization (PPO) \cite{schulman2017proximal} with training over 100 million steps. PPO is a policy gradient method that optimizes a neural network policy by alternating between sampling data through interactions with the environment and optimizing a surrogate objective function using stochastic gradient ascent. The approach includes a clip parameter that restricts policy updates to a trust region, preventing excessive policy changes that could lead to performance collapse.\\

In our experiments, we adopt the PPO implementation from Stable Baselines3 \cite{stablebaselines3}. We use a two-hidden-layer architecture (256 units per layer) with SiLU\footnote{SiLU (Sigmoid Linear Unit) or Swish function: $f(x) = x \cdot \sigma(x)$, where $\sigma(x) = \frac{1}{1 + e^{-x}}$ is the sigmoid function.} activations for both the policy and value function networks. More details on the hyperparameters used for the PPO model are provided in Table \ref{table:ppo_params}.\\

\begin{table}[h]
\centering
\begin{tabular}{|l|p{6cm}|l|}
\hline
\textbf{Parameter Name} & \textbf{Definition} & \textbf{Value} \\
\hline
policy\_kwargs & Neural network architecture specifications & \begin{tabular}[c]{@{}l@{}} pi=[256, 256] \\ vf=[256, 256] \\ activation\_fn=SiLU \end{tabular}\\ 
\hline
learning\_rate & Learning rate for the optimizer & 0.001 \\
\hline
gamma (RL) & Discount factor for future rewards & 0.99 \\
\hline
ent\_coef & Entropy coefficient for exploration & 0.01 \\
\hline
vf\_coef & Value function coefficient & 0.5 \\
\hline
max\_grad\_norm & Maximum value for gradient clipping & 1.0 \\
\hline
batch\_size & Number of samples in each optimization batch & 4000 \\
\hline 
\end{tabular}
\caption{PPO hyperparameters used in the experiments.}
\label{table:ppo_params}
\end{table}

\begin{figure}[h!]
  \centering
  \includegraphics[width=0.8\linewidth, trim={0 0 0 0}, clip]{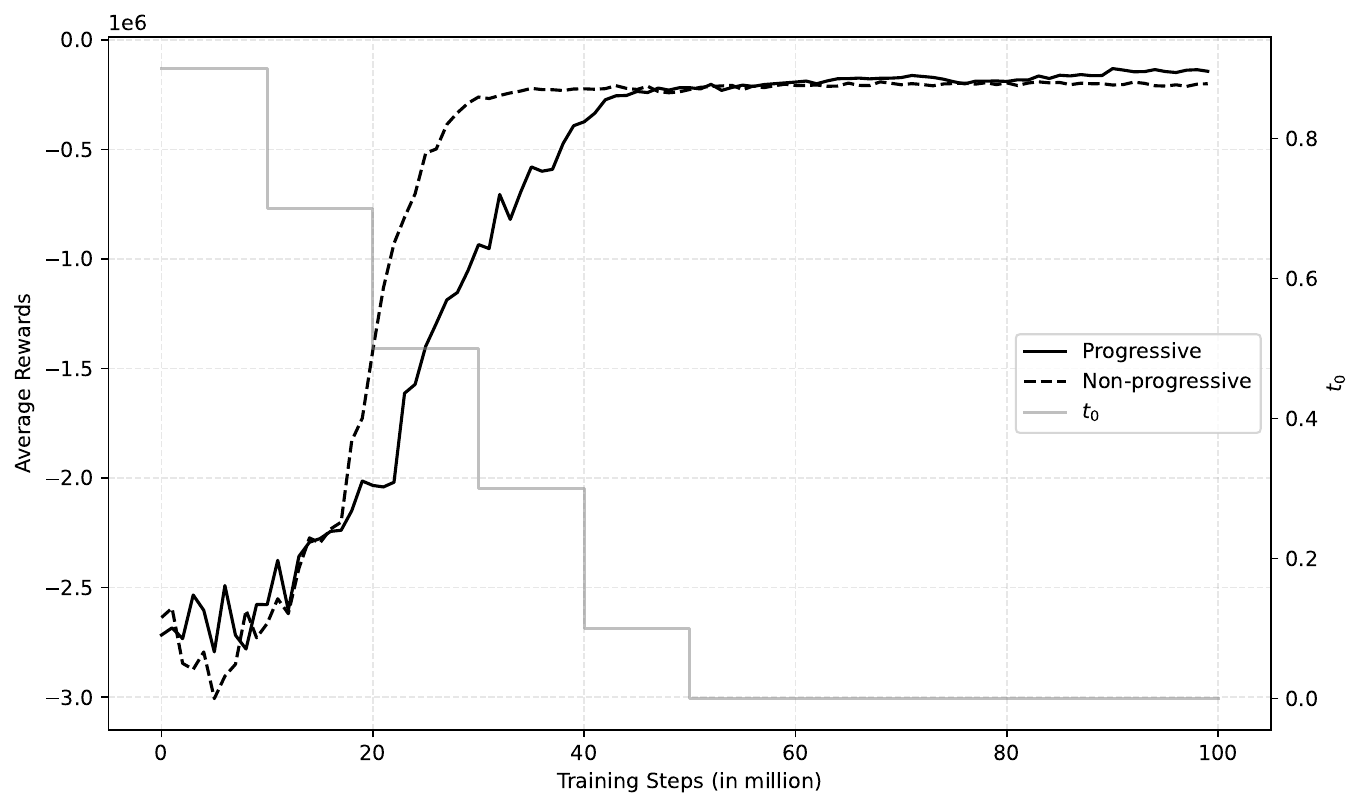}
\caption{Comparison of training approaches showing the average episode reward curves during the training of RL agents using the progressive learning method (solid line) and the non-progressive method (dashed line).}
  \label{fig:RL_comparison_progressive}
\end{figure}

\begin{figure}[h!]
  \centering
  \begin{minipage}{0.48\textwidth}
    \centering
    \includegraphics[width=\textwidth]{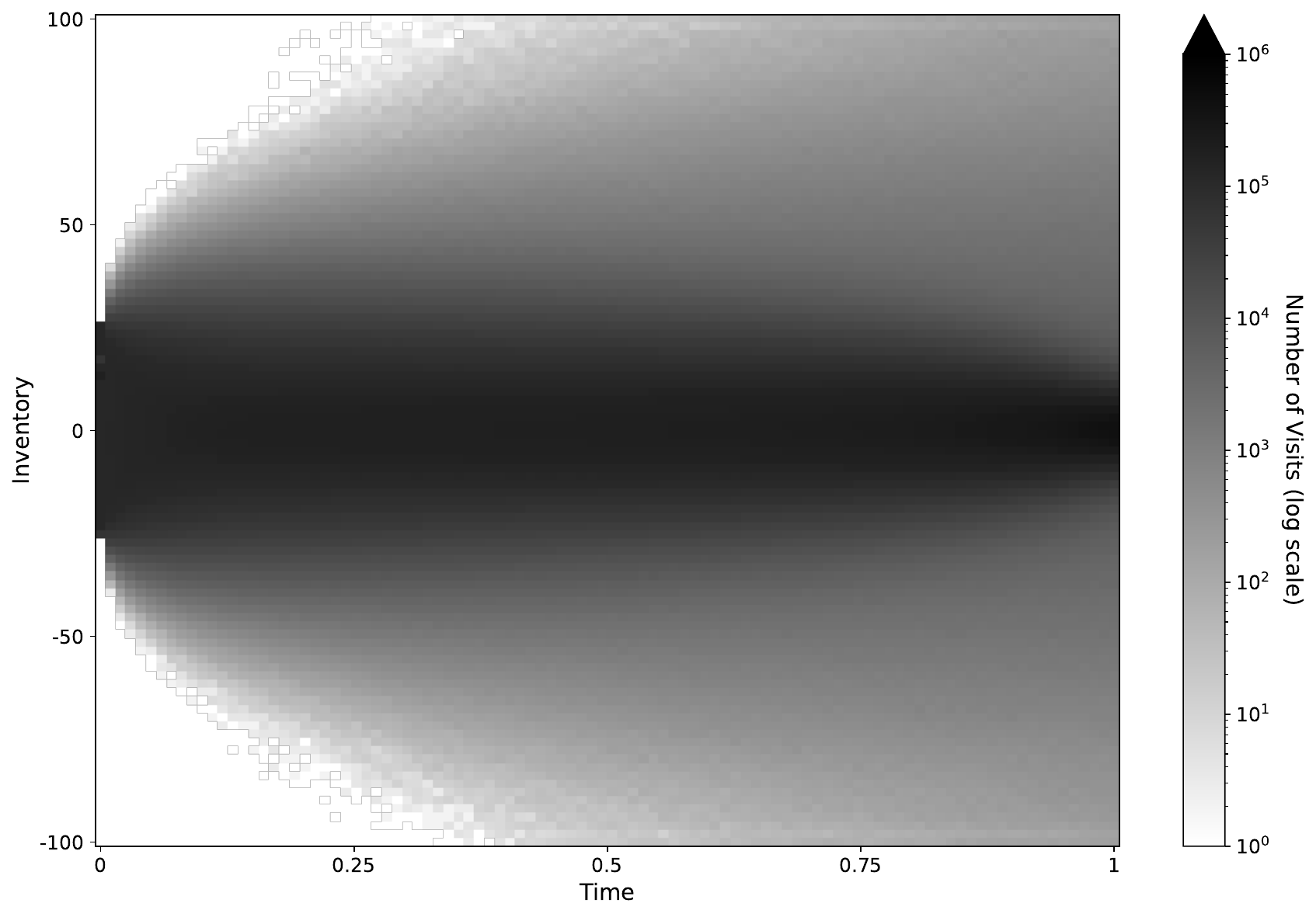}
  \end{minipage}
   \hfill
  \begin{minipage}{0.48\textwidth}
    \centering
    \includegraphics[width=\textwidth]{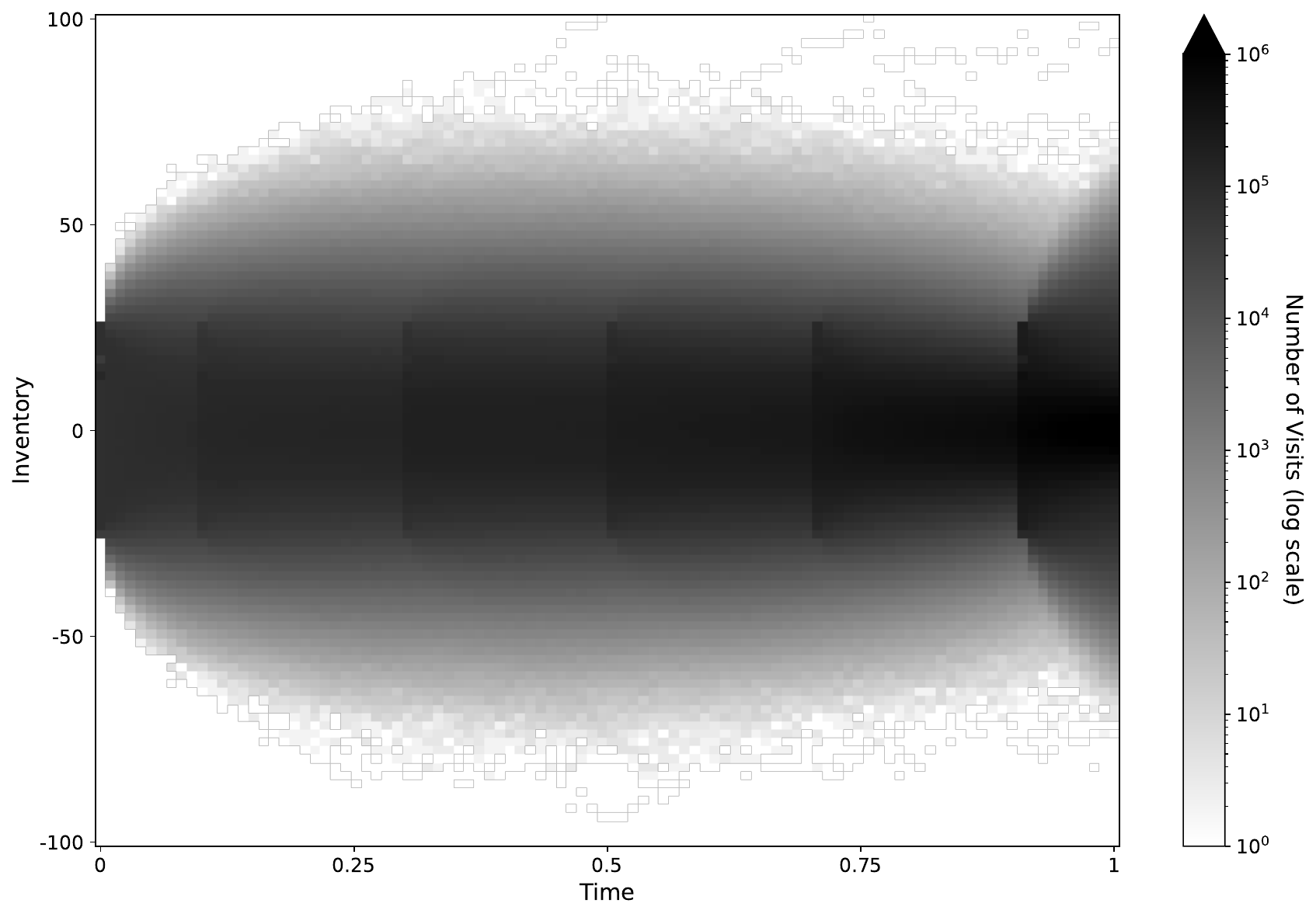}
  \end{minipage}
  \caption{Heatmaps of states visited during training: non-progressive training on full episodes (left) versus progressive training approach (right).}
  \label{fig:RL_comparison_progressive_states}
\end{figure}

Because the problem involves a time interval $[0,T]$, we implemented a progressive training approach. This approach initially trains the model on environments with states close to the terminal time and gradually extends the time horizon until covering the complete episode (from $t = 0$ to $t = T$).
Figure \ref{fig:RL_comparison_progressive} illustrates the performance difference between two training approaches: (1) training directly on complete episodes (non-progressive) and (2) using the progressive training method. In the progressive approach, the model is trained sequentially on environments with initial times $t_0 \in [0.9, 0.7, 0.5, 0.3, 0.1]$, allocating 10 million steps for each interval, followed by 50 million steps on complete episodes ($t_0 = 0$). It shows how the non-progressive approach may quickly converge to a sub-optimal solution, mainly due to the underexploration of some states. In fact, Figure \ref{fig:RL_comparison_progressive_states} presents the state space exploration patterns of both approaches. The progressive learning method exhibits more efficient exploration, concentrating on strategically important state regions (specifically, inventory levels between -35 and 35 lots). In contrast, the non-progressive approach shows suboptimal exploration patterns, particularly evident in states near $t = 1$.\\

To assess and compare the performance of different execution strategies, we establish a naive rule-based benchmark strategy and apply both reinforcement learning and PDE-based methods. The rule-based benchmark follows a simple inventory control rule: it trades at a constant speed of 100 lots per day whenever the inventory level (in absolute value) exceeds a predefined threshold of 5 lots, and does not trade otherwise. The RL-based approach aims to learn an optimal adaptive execution strategy by interacting with the market environment, dynamically adjusting its trading behavior based on real-time observations. Meanwhile, the PDE-based approach formulates the optimal execution problem using the Hamilton-Jacobi-Bellman equation presented in Eq.~(\ref{PDE_theta_B}), deriving an optimal control strategy through numerical methods. \\

For our numerical experiments, we consider three cases: two special cases where either $\psi$ or $\eta$ is set to zero, and a general case where both parameters are nonzero.

\subsubsection{Case 1: $\psi = 0$}

For the case where $\psi = 0$, corresponding to purely quadratic execution costs ($L(v) = \eta v^{2}$), the HJB equation admits a closed-form solution, and the corresponding optimal control is given in Eq.~(\ref{PDE_psi_0_solution_control}). This solution describes an optimal policy that is linearly dependent on inventory, with a slope that evolves as a function of the remaining time until time $T$. As time approaches~$T$, the absolute value of the slope increases rapidly, indicating that the optimal strategy is to gradually unwind the inventory early on and accelerate execution as the end of the trading period nears. 

\begin{figure}[h!]
    \centering
    \begin{minipage}{0.49\textwidth}
        \centering
        \includegraphics[width=\linewidth]{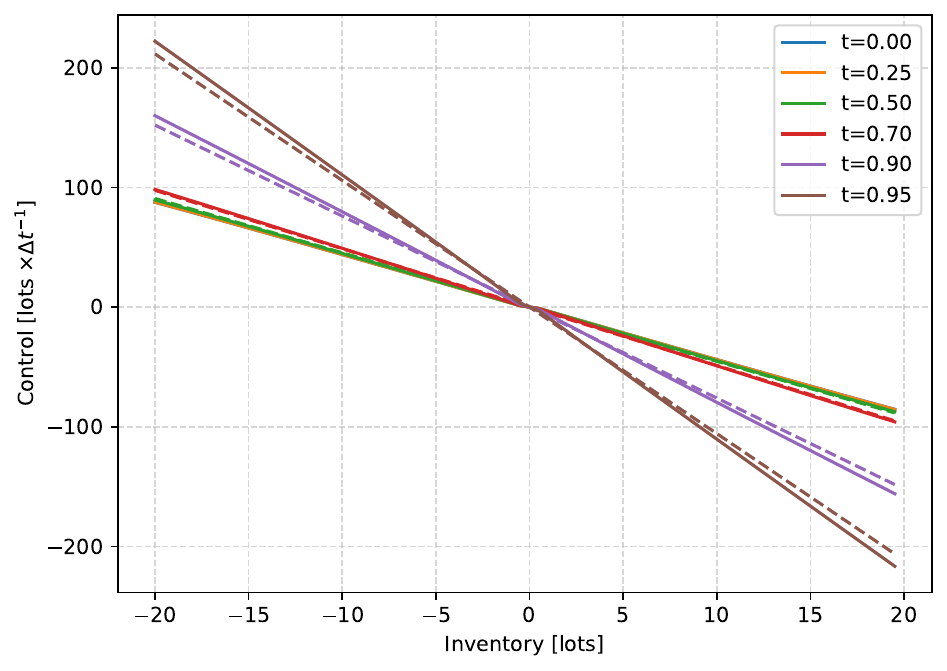}
    \end{minipage}
    \hfill
    \begin{minipage}{0.49\textwidth}
        \centering
        \includegraphics[width=\linewidth]{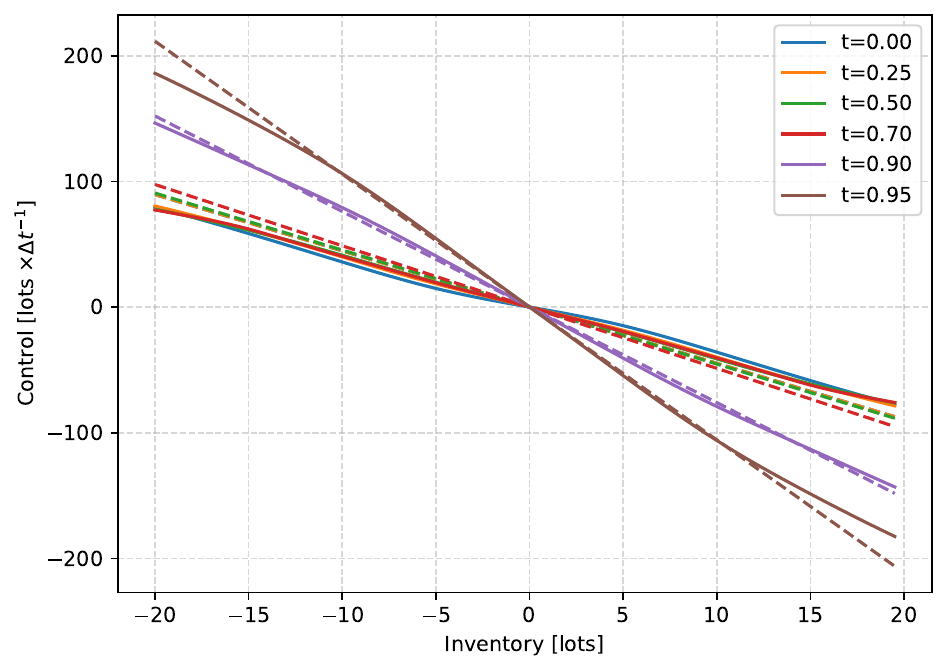}
    \end{minipage}
    \caption{Comparison of PDE-based optimal control (left) and RL-derived optimal policy (right) (solid lines) against the closed-form solution (dashed lines) at different time steps.}
    \label{fig:RL_psi_0_combined}
\end{figure}

Figure~\ref{fig:RL_psi_0_combined} compares the optimal policy obtained via the PDE numerical solution (on the left panel) and RL (on the right panel) against this theoretical benchmark. We see that both the RL-based and the PDE-based policies align well with the theoretical solution. However, for situations corresponding to large inventories, the approximations are not perfect. This is because these states are visited less frequently in the learning process in the case of RL, especially at the end of the training when the model has learned to avoid holding large inventories near time $T$. In the case of the PDE-based approach, this is due to the Neumann conditions at the boundaries.\\

\begin{figure}[h!]
    \centering
    \includegraphics[width=0.62\textwidth]{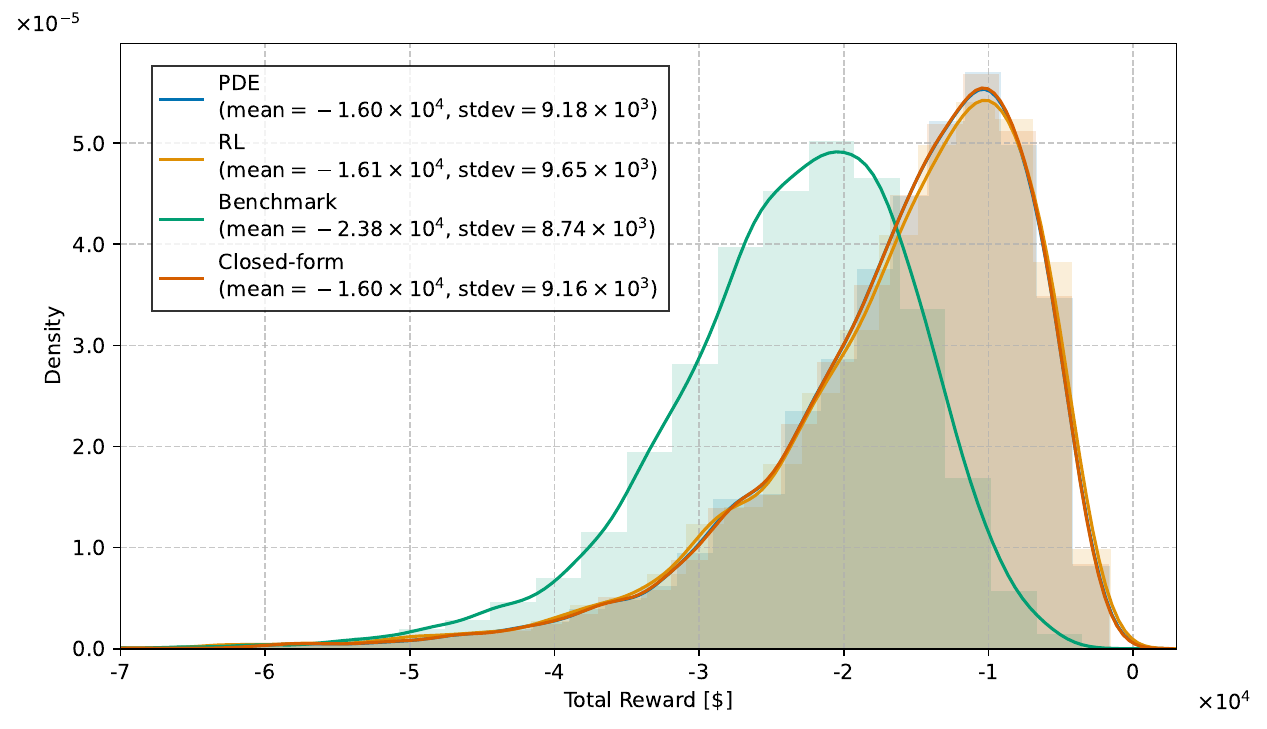}
\caption{Histograms and density curves of total rewards obtained from 10,000 out-of-sample simulated episodes for the case $\psi = 0$. The density curves are estimated using Kernel Density Estimation (KDE) with a Gaussian kernel. The PDE-based solution produces a density curve that is very similar to the closed-form solution, making them nearly indistinguishable in the plot.}
    \label{fig:RL_psi_0_dist}
\end{figure}

Figure~\ref{fig:RL_psi_0_dist} compares the total rewards achieved by the different strategies (the strategy associated with the theoretical solution is labeled as \textit{Closed-form} in the figure). We clearly see that the naive rule-based benchmark is outperformed by both the RL and PDE-based solutions, which exhibit performance very close to the optimal closed-form solution.

\subsubsection{Case 2: $\eta = 0$}

\begin{figure}[h!]
    \centering
    \includegraphics[width=0.9\textwidth]{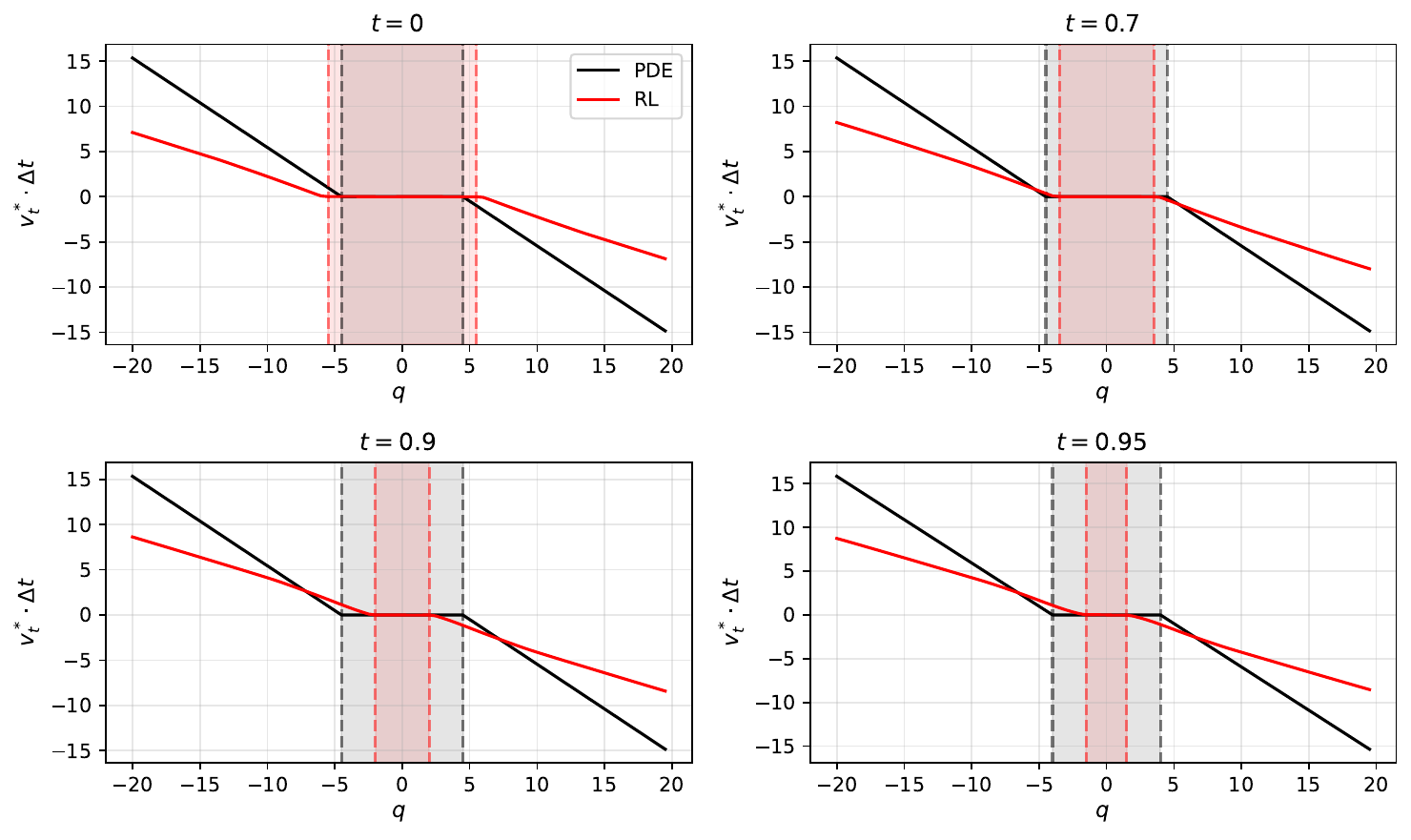}
\caption{RL-based and PDE-based policies for different times. Shaded regions indicate no-trading zones as identified by each strategy.}
    \label{fig:RL_eta_0_actions}
\end{figure}

In the case when $\eta = 0$, where execution costs are purely linear ($L(v) = \frac{\psi}{2} |v|$), there is no closed-form solution, so we rely on a comparison between the RL policy and the policy derived from the numerical solution of the HJB equation. Figure~\ref{fig:RL_eta_0_actions} shows the policies corresponding to the two numerical methods. \\

A key observation in this case is the structure of the optimal control, which can be divided into two distinct zones. There exists an interval where it is optimal to remain inactive, executing no trade (full internalisation), while outside of this interval externalisation is optimal. Additionally, as the terminal time approaches, the no-execution zone shrinks to avoid terminal inventory penalties. For the PDE-based strategy, externalisation consists in trading a quantity that sends the inventory back to the frontier of the pure-internalisation interval. The reinforcement learning agent successfully captures the policy structure, with no execution within a given interval and an execution speed that increases with inventory outside of this interval. However, the frontier between internalisation and externalisation varies more over time and the RL agent does not bring the inventory back to the frontier in one step but adopts a less aggressive approach.\\

\begin{figure}[h!]
    \centering
    \includegraphics[width=0.62\textwidth]{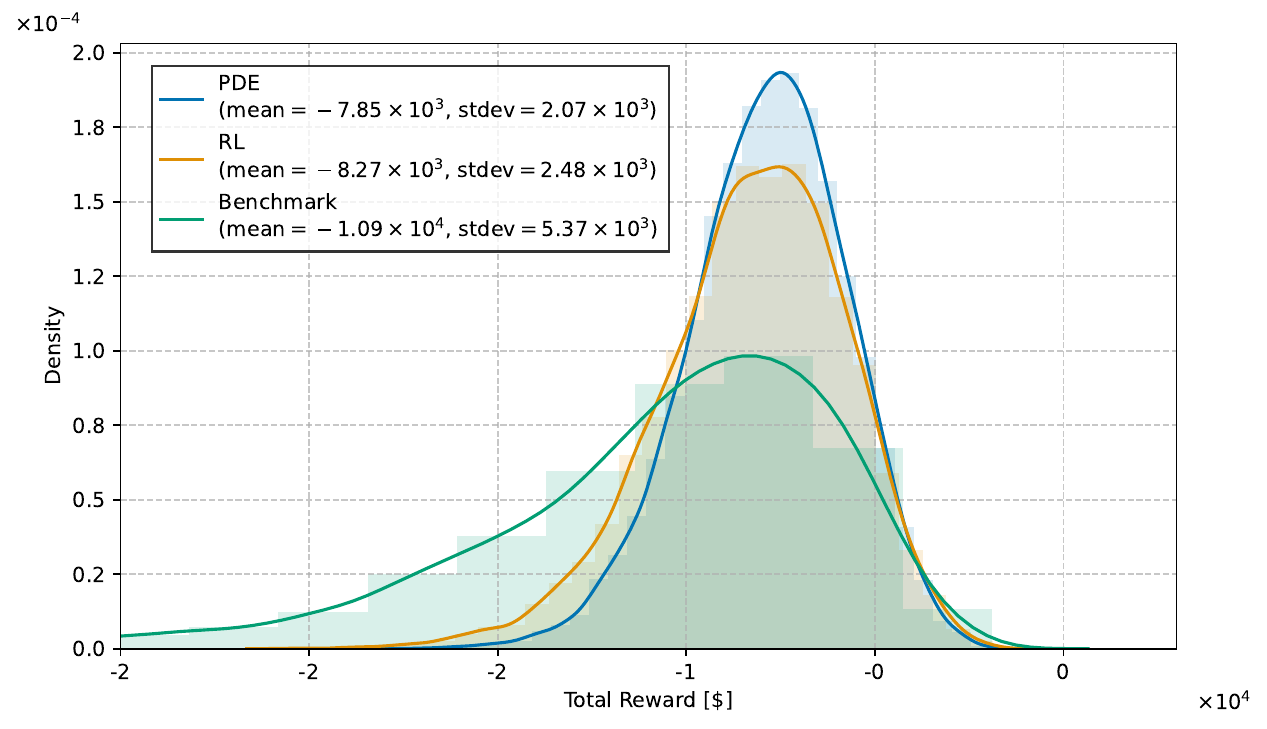}
\caption{Histograms and density curves of total rewards obtained from 10,000 out-of-sample simulated episodes for the case $\eta = 0$. The density curves are estimated using KDE with a Gaussian kernel.}
    \label{fig:RL_eta_0_dist}
\end{figure}

Figure~\ref{fig:RL_eta_0_dist} compares the performance of the different strategies in an out-of-sample simulated environment. Both the HJB numerical solution and the RL policy significantly outperform the rule-based benchmark. On average, the PDE-based solution achieves a slight but consistent outperformance compared to the RL-based policy. However, this difference may be due to local maxima or flat regions in the objective function, resulting in multiple near-optimal solutions with similar value estimates. The latter may also explain the notable difference between the learned RL policy and the policy resulting from the numerical scheme, presented in Figure~\ref{fig:RL_eta_0_actions}.\\

\subsubsection{Case 3: $\psi >0$ and $ \eta > 0$}

\begin{figure}[h!]
    \centering
    \includegraphics[width=0.9\textwidth]{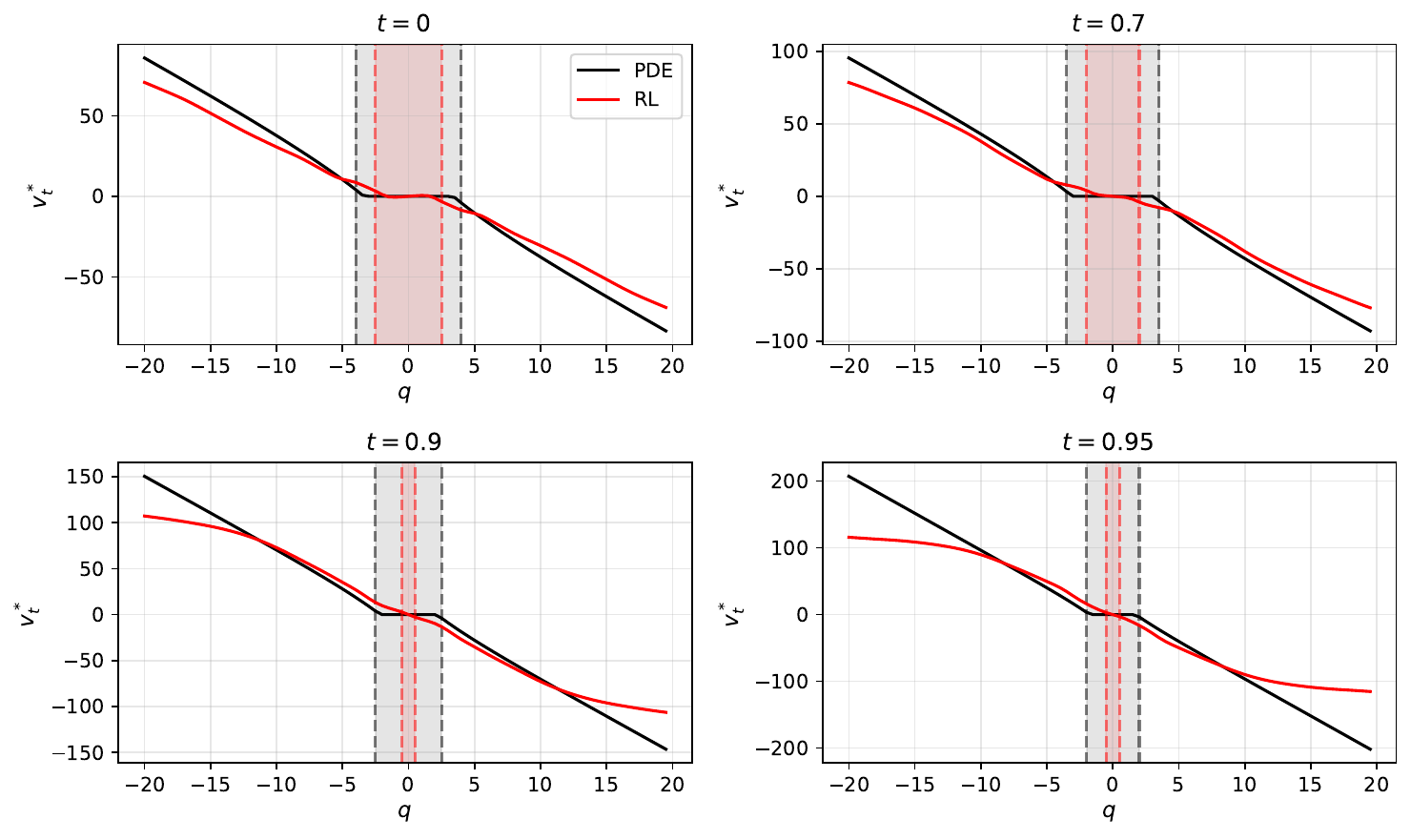}
\caption{RL-based and PDE-based policies for different times. Shaded regions indicate no-trading zones as identified by each strategy.}
    \label{fig:RL_general_actions}
\end{figure}

We now consider the case where the execution costs are composed of both linear and quadratic terms, i.e. $L(v) = \frac{\psi}{2} |v| + \eta v^{2}$. Figure~\ref{fig:RL_general_actions} compares the RL-based and PDE-based policies for different times and inventories.  Because $\psi > 0$, the PDE-based solution exhibits intervals where the control is $0$ (internalisation) when the inventory is small (in absolute value), and the pure-internalisation interval gets smaller as we approach the terminal time. The RL-based policy exhibits similar characteristics but appears to be less aggressive for larger inventory values as in the pure quadratic case.\\

\begin{figure}[h!]
    \centering
    \includegraphics[width=0.62\textwidth]{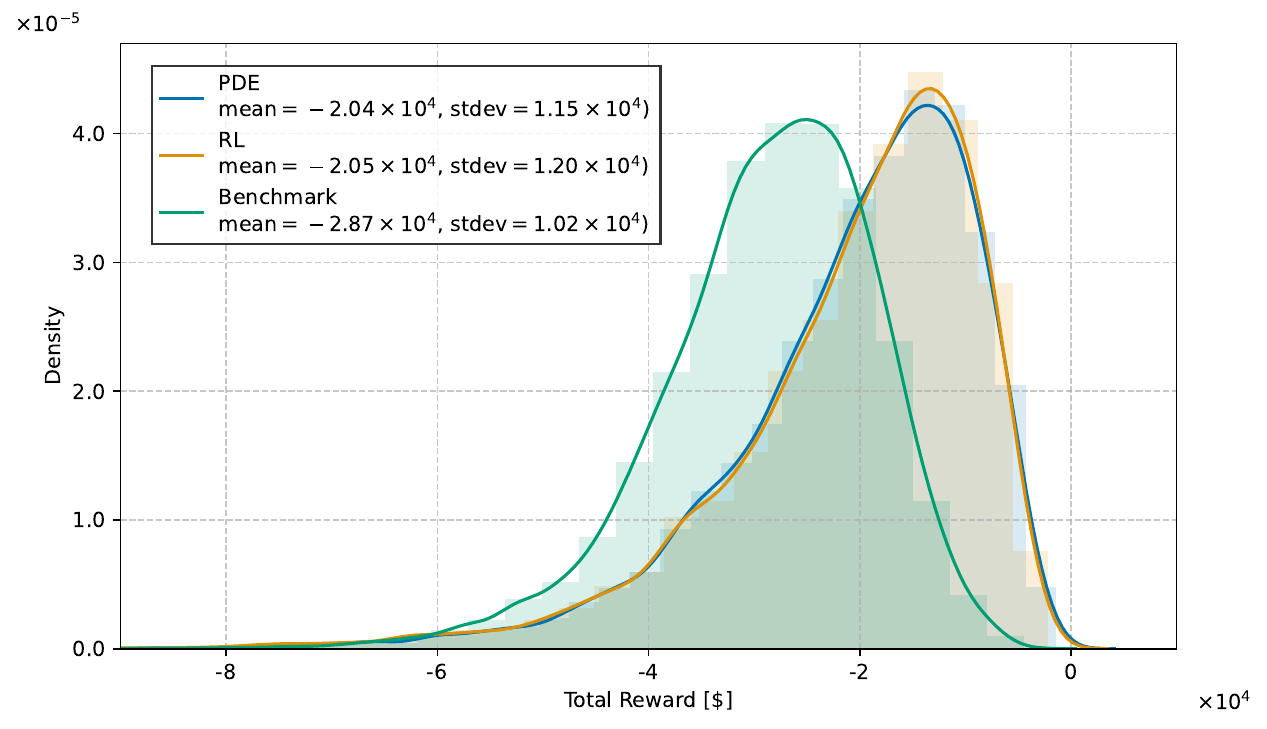}
\caption{Histograms and density curves of total rewards obtained from 10,000 out-of-sample simulated episodes for the general case. The density curves are estimated using KDE with a Gaussian kernel.}
    \label{fig:RL_general_dist}
\end{figure}

Performance-wise, both approaches significantly outperform the naive benchmark policy, as illustrated in Figure~\ref{fig:RL_general_dist}. The RL-based policy achieves an average accumulated reward within 0.5\% of the PDE-base one.\footnote{The hypothesis of equal means is not rejected based on a two-sample t-test ($p$-value = 0.07). Similarly, the Mann-Whitney U test does not reject the hypothesis that both distributions are identical ($p$-value = 0.87). Both tests use a significance threshold of 5\%.} While the RL-based policy leads to slightly higher variance, it successfully learns a near-optimal control strategy.\\

\section{Conclusion}

In this paper we studied the problem of optimally managing a stochastic order flow, taking into account market impact and execution costs as in models \textit{à la} Almgren-Chriss. We proposed the use of an RL-based approach as an alternative to a more classical PDE-based approach in order to determine optimal trading policies.\\

According to our results, our RL method converges to near-optimal solutions successfully, resulting in performance almost similar to the PDE-based solution, or exact solutions in contexts where they are known. A major benefit of RL is that no boundary conditions are needed, unlike with PDEs.\\

As RL generalizes well to higher-dimensional problems where traditional PDE solvers suffer from the curse of dimensionality and can handle more complex market dynamics, it constitutes a promising approach for solving real-world execution problems.

\section*{Context and Acknowledgements}

This paper is partly based on previous unpublished research conducted by Philippe Bergault and Olivier Guéant, supported by the Research Initiative ``Nouveaux traitements pour les données lacunaires issues des activités de crédit,'' financed by BNP Paribas under the aegis of the Europlace Institute of Finance. Hamza Bodor benefited from a ``Bourse CIFRE'' financed by BNP Paribas and contributed to reviving the topic with his expertise in reinforcement learning.\\

The authors would like to express their gratitude for the continuous support provided by BNP Paribas. Nonetheless, the content of this article does not, in any manner, reflect the practices or policies of BNP Paribas.\\

\bibliographystyle{plain}

\end{document}